\documentclass[11pt]{article}
\usepackage{epsfig}
\usepackage{amssymb}
\usepackage{amsmath}

\newcommand{\ud}{\mathrm{d}}

\newcommand{\be}{\begin{equation}}
\newcommand{\ee}{\end{equation}}
\newcommand{\bea}{\begin{eqnarray}}
\newcommand{\eea}{\end{eqnarray}}
\newcommand{\bml}{\begin{mathletters}}
\newcommand{\eml}{\end{mathletters}}

\begin{document}

\title{\LARGE \bf Perfect Fluid Spherically-Symmetric Solutions\\ in Massive Gravity}
 \author{
  \large  Y. Brihaye$^a$ {\em and}$\:$  Y. Verbin$^b$
  \thanks{{Electronic addresses: yves.brihaye@umh.ac.be; verbin@openu.ac.il} } }
 \date{ }
   \maketitle
    \centerline{$^a$ \em Faculte des Sciences, Universite de Mons, }
     \centerline{\em 7000 Mons, Belgium}
     \vskip 0.4cm
     \centerline{$^b$ \em Department of Natural Sciences, The Open University
   of Israel,}
   \centerline{\em Raanana 43107, Israel}

\date{\ }

\maketitle
\thispagestyle{empty}
\begin{abstract}
We study spherically-symmetric solutions in Massive Gravity generated by matter
sources with polytropic equation of state. We concentrate in the non-perturbative regime
where the mass term non-linearities are important, and present the main features of the solutions.

\end{abstract}
\maketitle
\medskip \medskip

\section{Introduction }\label{Intr}
\setcounter{equation}{0}

Since the discovery of the late time cosmic acceleration in the 1990's \cite{RiessEtAl1998,PerlmutterEtAl1998},
a considerable amount of effort has been invested in order to understand its origin.
The simplest explanation is the old idea
of the cosmological constant which requires extreme fine-tuning in order to fit its value with the
present matter density on one hand, and with possible microscopic mechanisms for its generation on
the other hand \cite{Weinberg1989}. This difficulty which is known as the {\it cosmological constant
problem} motivates the search of alternative explanations for the cosmic acceleration,
which usually are taken as one of the following two approaches:

The first assumes some unknown form of matter with an equation of state
similar to that of a cosmological constant: $P\simeq -\rho$. This {\it dark energy} is assumed to
dominate the universe at recent times (starting around $z\simeq 0.5$) thus changing its expansion
rate from the decelerating mode of matter-dominated universe to the accelerating one. Quintessence
models \cite{Quint1,Quint2} are one popular example of this approach.

The second  approach suggests a modification of the theory of General Relativity (GR) in a way that produces
deviations in a cosmological scale. The hope is that the modifications will have such an effect which will
enable to explain the cosmic acceleration without invoking dark energy.

These approaches may also be relevant in tackling the Dark Matter problem and the puzzle of coincidence,
that is, that these two components, dark energy and dark matter have a similar weight in the present
cosmological epoch.

Massive Gravity or Massive General Relativity (MGR) \cite{Hinterbichler} is a modification of GR
originated from  the very natural question: what is the way to give the graviton a mass (or a finite range)?
This question was partially answered back in the 1930's by Fierz and Pauli \cite{FierzPauli1,FierzPauli2}
who wrote down a theory of a massive spin 2 field in Minkowski background.
However, the generalization to the full self-gravitating case was proved to be notoriously difficult.
Some of the hurdles were: \vspace {-0.6cm}
\begin{itemize}
\item the van Dam--Veltman--Zakharov (vDVZ) discontinuity
\cite{vanDamVeltman,Zakharov1970}  according to which gravitational attraction in MGR does not reproduce
 that of GR at the massless graviton limit.  \vspace {-0.6cm}
\item the discovery by Boulware and Deser \cite{BoulwareDeser1,BoulwareDeser2} (BD)
that a generic extension of this theory to curved backgrounds contains a ghost degree of freedom in addition to
the five of the massive spin 2 graviton.
\end{itemize}

A possible way out of the vDVZ discontinuity is known as the Vainshtein mechanism \cite{Vainshtein}.
 Vainshtein realized that in such a theory one cannot rely on weak field results (as
vDVZ did), since the non-linearities become stronger as the mass (say $m$) of the graviton decreases.
This phenomenon is quantified by the {\it Vainshtein radius} associated with a localized massive source of
 mass $M$. This new length-scale $r_V$
 determines the range beyond which linear approximations are justified. The fact that $r_V$ increases indefinitely as
 $m$ decreases (see e.g.  Eq. (\ref{RVainsht}) below) implies immediately that as $m\rightarrow 0$, the linear theory cannot be trusted anywhere.
 This opens the possibility  that  nonlinear effects will cure the vDVZ discontinuity as is indeed verified
 by an explicit analysis  \cite{KoyamaEtAl1}.

However, due to the BD ghost, MGR was considered for a long time as inconsistent and/or unphysical.
Only very recently a solution was found  \cite{deRhamEtAl2011,HassanRosen}
and the theory became acceptable (see also  \cite{Hinterbichler} and references therein).

It is quite obvious that a  graviton mass term in a generally covariant theory of gravity cannot contain
derivatives of $g_{\mu\nu}$ on one hand, and cannot make use of the only two possible ``derivative-free''
quantities $tr(g^\mu _\nu)$ and $det(g_{\mu\nu})$ on the other hand. The conclusion is that a mass term cannot
be based on the metric tensor alone.

A direct way to overcome this is to introduce an additional rank-2 tensor, say $H_{\mu\nu}$ which can be used
to construct together with $g_{\mu\nu}$ non-trivial scalars
\cite{deRhamEtAl2011,HassanRosen,ChamseddineMukh1,AlberteChamseddineMukh}.
This $H_{\mu\nu}$ may be regarded as an auxiliary metric and be parametrized
by four St\"{u}ckelberg fields $\varphi^{a}$ such that
$H_{\mu\nu}=\eta_{ab}\partial_\mu \varphi^{a}\partial_\nu \varphi^{b}$ where
$\eta_{ab}$ is the Minkowski metric. Next define its square root $J_{\mu}^{\nu}$ by
$J_\mu^\lambda J_\lambda^\nu =H_\mu^\nu$ and the tensor $K_\mu^\nu = \delta_\mu^\nu - J_\mu^\nu $.
It was shown \cite{deRhamEtAl2011,HassanRosen} that there are only three combinations, quadratic,
cubic and quartic that can eliminate the BD ghost in curved background and eventually
yield a ghost-free theory. These contributions are written in terms of various traces of this $K_\mu^\nu$
tensor, namely ${\cal K }_n = tr(K^n)$, or actually the following combinations:
\begin{eqnarray} \nonumber
{\cal M }_2 = ({\cal K }_1)^2-{\cal K }_2 \;\;  , \;\;\;\;
{\cal M }_3 = ({\cal K }_1)^3 -3{\cal K }_1{\cal K }_2 +2{\cal K }_3 \; ,\\
{\cal M }_4 = ({\cal K }_1)^4 -6({\cal K }_1)^2{\cal K }_2 +3({\cal K }_2)^2 +8{\cal K }_1{\cal K }_3- 6{\cal K }_4
\label{traces}
\end{eqnarray}
such that the most general gravitational mass term ${\cal M }$ is
\begin{eqnarray} \nonumber
{\cal M }=m^2 \left( \frac{\sigma_2}{2}{\cal M }_2 + \frac{\sigma_3}{6}{\cal M }_3 +
\frac{\sigma_4}{24}{\cal M }_4 \right )=
m^2 [ \sigma_2 \left( k_0 (k_1 + k_2 + k_3) + k_1 (k_2 + k_3) + k_2 k_3 \right )+ \\
\sigma_3 (k_0 k_1 k_2 + k_0 k_1 k_3 +k_0 k_2 k_3 +k_1 k_2 k_3 )+ \sigma_4 k_0 k_1 k_2 k_3 ] \;\;\;\;\;\;
\label{MassTerm}
\end{eqnarray}
where $\sigma_2$, $\sigma_3$ and $\sigma_4$ are arbitrary dimensionless coefficients and the last expression uses
the eigenvalues $k_\alpha$ of $K_\mu^\nu$. Actually, one of these $\sigma_i$s can be absorbed by a redefinition of
$m$, but we keep all three in order to allow each of them to vanish.

The MGR action is thus given by the following GR contribution ($\kappa = 8\pi G$) with the additional three-parameter
mass term (and matter Lagrangian):
\begin{equation}
S=\int \ud^4x\sqrt{|g|}\left[ \frac{1}{2\kappa}({\cal R }+2{\cal M })+{\cal L}_{matter}\right]
\label{MGRaction}
\end{equation}
This form of the mass term has the effect of eliminating the sixth propagating mode that
typically exists in such a theory, leaving us with the five required for a massive spin 2 field.

Although the original motivation for studying MGR was pure physics curiosity, it fits well to the arena of the
dark energy and dark matter problems. A massive graviton is bound to introduce large distance effects into
gravity and may have its say on the DM problem as well.

MGR is presently in an initial state with respect to these issues,
due to the harsh theoretical obstacles that were mentioned above.
Only recently some results about Cosmology within the MGR framework have been published
 \cite{ChamseddineVolkov2011,D'AmicoEtAl2011,GumrukcuogluEtAl2011,ComelliEtAl2011FRW}.

As other modified theories of gravity, MGR not only introduces deviations from GR at large distances,
but also modifies the gravitational fields around compact sources. These modifications must be consistent
with the stringent observational solar system tests, but still may be used in order to differentiate between
the various modified gravity theories.

A natural way to go is analyzing spherically-symmetric solution of MGR. Some preliminary work in this direction has been
done recently \cite{KoyamaEtAl1,Nieuwenhuizen2011,Gruzinov2011,ComelliEtAl2011Sph,Sjors+Mortsell} and here we take this direction
and study localized self-gravitating solutions in MGR where the matter source is a perfect fluid with a polytropic
equation of state.

Observations already limit the graviton mass to a tiny value of $10^{-33}$eV at most
 \cite{D'AmicoEtAl2011,GoldhaberNieto}.  However, in order to understand the impact of the graviton mass term
 we will concentrate in the non-perturbative regime where the mass term non-linearities are important. This means
 that we will deal with structures whose sizes are of the order of the graviton range.

\section{Spherically-Symmetric Solutions}  \label{spherical}
\setcounter{equation}{0}

For spherical symmetry we use the most general \emph{diagonal} line element
\begin{equation}
ds^2=a^2( \textsl{r}) dt^2-v^2(\textsl{r}) d\textsl{r}^2 - \textsl{r}^2
c^2(\textsl{r})(d\theta^2+\sin^2(\theta)d\varphi^2)
\label{metric0}
\end{equation}
and the unitary gauge where
\begin{equation}
\eta_{ab}\partial_\mu \phi^{a}\partial_\nu \phi^{b}dx^\mu dx^\nu= dt^2-d\textsl{r}^2 -
\textsl{r}^2(d\theta^2+\sin^2(\theta)d\varphi^2)
\label{UnitaryG}.
\end{equation}
It is well-known \cite{KoyamaEtAl1} that this way leads to asymptotically flat solutions which is the branch we are
interested in. Note that we cannot redefine the radial coordinate to get rid of $c(\textsl{r})$ (as is done in GR) without breaking the
Lorenz symmetry of the unitary gauge. We will have therefore to solve here for three components of the metric tensor.

In these coordinates the tensor $K_\mu^\nu$ will be diagonal with components
\begin{equation}
k_0=1-1/a \;\; , \;\; k_1=1-1/v , \;\; k_2=k_3=1-1/c
\label{diagK0}.
\end{equation}
We can however perform a coordinate transformation $\textsl{r} (r)$ such that the new line element will be
\begin{equation}
ds^2=a(r)^2 dt^2-dr^2/b(r)^2 - r^2(d\theta^2+\sin^2(\theta)d\varphi^2)
\label{metric}
\end{equation}
so the $K_\mu^\nu$ components become now
\begin{equation}
k_0=1-1/a \;\; , \;\; k_1=1-b(r/c)' , \;\; k_2=k_3=1-1/c
\label{diagK}.
\end{equation}
The field equations obtained from the action (\ref{MGRaction}) have the general form
\begin{equation}
G_{\mu\nu} + S_{\mu\nu} + \kappa T_{\mu\nu} =0
\label{FEq}
\end{equation}
where $S_{\mu\nu}$ is the contribution from the mass term. In the coordinate system where the metric is given by
Eq. (\ref{metric}) the components of Einstein tensor are:
\begin{eqnarray} \nonumber
G^0_0 &=& \frac{2 b b'}{r} +\frac{b^2-1}{r^2} \\
G^r_r &=& \frac{2 b^2 a'}{r a} + \frac{b^2-1}{r^2} \\ \nonumber
G^\theta_\theta = G^\varphi_\varphi &=&
b^2 \left(\frac{a''}{a}+\frac{a' b'}{ab}+\frac{a'}{r a}+\frac{b'}{r b}\right)
\label{EinstT}
\end{eqnarray}
and the mass term contribution is
\begin{eqnarray} \nonumber
S^0_0 &=& -\mu _3 +\frac{2 \mu _2}{c}-\frac{\mu _1}{c^2}+\left(\mu _2-\frac{2 \mu _1}{c}
+\frac{\mu _0}{c^2}\right) \frac{b\left(c-r
   c'\right)}{c^2}  \\
S^r_r &=& -\mu _3 +\frac{2 \mu _2}{c}-\frac{\mu _1}{c^2}+\frac{1}{a}\left(\mu _2-\frac{2 \mu _1}{c}
+\frac{\mu _0}{c^2}\right)\\ \nonumber
S^\theta_\theta = S^\varphi_\varphi &=& -\mu_3 +\mu _2 \left(\frac{1}{a}+\frac{1}{c}\right) -\frac{\mu _1}{a c}
+\left(\mu _2 -\mu _1 \left(\frac{1}{a}+\frac{1}{c}\right)+\frac{\mu _0}{ac}\right)
\frac{b \left(c-r c'\right)}{c^2}
\label{MassT}
\end{eqnarray}
where we denote $\mu_1 = m^2 (\sigma_2 + 2 \sigma_3 + \sigma_4 ) $,
$\mu_2 =  m^2 (3\sigma_2 + 3 \sigma_3 + \sigma_4 ) $,
$\mu_3 =  m^2 (6\sigma_2 + 4 \sigma_3 + \sigma_4 )$, $\mu_0 =  m^2 (\sigma_3 + \sigma_4)$.

The components of the matter energy-momentum tensor may be added to the field equations as
$T_\mu^\nu = diag(\rho (r), -P_r (r), -P_\bot (r), -P_\bot (r))$. These components satisfy a continuity
equation
\begin{equation}
P_r '+ (\rho  + P_r)\frac{a'}{a}+\frac{2 (P_r -P_\bot)}{r} = 0
\label{ContMatter}
\end{equation}
which is equivalent to the matter field equation if the matter is described by a Lagrangian. Otherwise,
an additional equation of state is required.

Since the matter satisfies the continuity equation separately, the mass term $S_{\mu\nu}$ should be also
separately conserved. Explicitly we get
\begin{equation}
\left(\mu _2-\frac{2 \mu _1}{c}+\frac{\mu _0}{c^2}\right)\frac{r a'}{a}+\frac{2 (b-1)}{b}
\left(\mu_2-\mu _1 \left(\frac{1}{a}+\frac{1}{c}\right)+\frac{\mu _0}{ac}\right) = 0
\label{ContMGR}
\end{equation}
which is of course not an independent equation but can be useful for the analysis.

Note the four (or actually one) functions which appear repeatedly in the equations:
\begin{equation}
f_i (a,c)=\mu_i -\mu_{i-1} \left(\frac{1}{a}+\frac{1}{c}\right)+\frac{\mu_{i-2}}{ac} \,\,\,\, , i=2,3
\label{f-fctn}
\end{equation}
or more explicitly:
\begin{eqnarray} \nonumber
f_2 (a,c)=\mu _2 -\mu _1 \left(\frac{1}{a}+\frac{1}{c}\right)+\frac{\mu _0}{ac}
\,\,\,\, ; \,\,\,\, g_2 (c)=f_2 (c,c) \\
f_3 (a,c)=\mu _3 -\mu _2 \left(\frac{1}{a}+\frac{1}{c}\right)+\frac{\mu _1}{ac}
\,\,\,\, ; \,\,\,\, g_3 (c)=f_3 (c,c)
\label{fg-funct}
\end{eqnarray}
 so the field equations will get the following compact form:
\begin{eqnarray} \nonumber
\frac{2 b b'}{r} +\frac{b^2-1}{r^2} -g_3 (c)+g_2 (c) \frac{b\left(c-rc'\right)}{c^2} +\kappa\rho= 0\\ \nonumber
\frac{2 b^2 a'}{r a} + \frac{b^2-1}{r^2}
-g_3 (c)+\frac{1}{a} g_2 (c) -\kappa P_r= 0\\ \nonumber
b^2 \left(\frac{a''}{a}+\frac{a' b'}{ab}+\frac{a'}{r a}+\frac{b'}{r b}\right)
-f_3 (a,c) +f_2 (a,c)\frac{b \left(c-r c'\right)}{c^2} -\kappa P_\bot= 0\\
g_2 (c)\frac{r a'}{a}+\frac{2 (b-1)}{b} f_2 (a,c) = 0
\label{EqsCompact}
\end{eqnarray}
and in expanded form:
\begin{eqnarray} \nonumber
\frac{2 b b'}{r} +\frac{b^2-1}{r^2} -\mu _3 +\frac{2 \mu _2}{c}-\frac{\mu _1}{c^2}+\left (\mu _2-\frac{2 \mu _1}{c}
+\frac{\mu _0}{c^2}\right) \frac{b\left(c-rc'\right)}{c^2} +\kappa\rho= 0\\ \nonumber
\frac{2 b^2 a'}{r a} + \frac{b^2-1}{r^2}
-\mu _3 +\frac{2 \mu _2}{c}-\frac{\mu _1}{c^2}+\frac{1}{a} \left(\mu _2-\frac{2 \mu _1}{c}
+\frac{\mu _0}{c^2}\right) -\kappa P_r= 0\\  \nonumber
b^2 \left(\frac{a''}{a}+\frac{a' b'}{ab}+\frac{a'}{r a}+\frac{b'}{r b}\right)
-\mu _3 +\mu _2 \left(\frac{1}{a}+\frac{1}{c}\right)-\frac{\mu _1}{ac} +   \hspace{3cm}\\  \nonumber
\left (\mu _2 -\mu _1 \left(\frac{1}{a}+\frac{1}{c}\right)+\frac{\mu _0}{ac}\right)\frac{b \left(c-r c'\right)}{c^2}
-\kappa P_\bot= 0\\
\left(\mu _2-\frac{2 \mu _1}{c}
+\frac{\mu _0}{c^2}\right)\frac{r a'}{a}+
\frac{2 (b-1)}{b} \left ( \mu _2 -\mu _1 \left(\frac{1}{a}+\frac{1}{c}\right)+\frac{\mu _0}{ac}\right) = 0
\label{EqsNonCompact}
\end{eqnarray}
We may eliminate $b(r)$ using the last of Eq. (\ref{EqsNonCompact}) or alternatively, using also the second
equation of (\ref{EqsNonCompact}), express $a(r)$ in terms of
$b(r)$, $c(r)$ and $P_r (r)$ and solve for these three variables.  This leads to an algebraic expression
    \begin{equation}
   \label{a_expression}
    a(r) = Y(r,b,c,P_r)/Z(r,b,c,P_r)
    \end{equation}
 where $Y$ and $Z$ are polynomials in their arguments.
  The equations (\ref{EqsNonCompact})  then reduce to a
  system of two first order equations for  $b(r)$ and $c(r)$. This is the approach that
  we followed to construct solutions. There exist, to our knowledge, no explicit solutions to the system of
  equations above (even for the vacuum equations), so the way to proceed was to turn to numerical methods and to make
  simple assumptions about the matter source.

 For the vacuum solutions we have however the following asymptotic behavior (for $mr>>1$) \cite{Hinterbichler,Sjors+Mortsell}:
 \begin{eqnarray} \nonumber
\label{yukawa}
      a(r) = 1 - \frac{4GM}{3} \frac{e^{-mr}}{r} \ \ , \ \ \ \ b(r) = 1 - \frac{2GM}{3} \frac{e^{-mr}(1+m r)}{r} \ \ ,\\
          c(r) = 1 + \frac{2GMe^{-mr}}{3r} \left(1 + \frac{1}{mr} + \frac{1}{m^2 r^2}\right)
\end{eqnarray}
where the integration constants are expressed in terms of the (``gravitational'') mass $M$ of the source
obtained from a weak field correspondence. This will be our way to identify the  mass of
the solutions we will find.

\section{Polytrope Solutions} \label{poly}
\setcounter{equation}{0}

We will concentrate in this work in localized asymptotically flat solutions produced by a perfect fluid. Asymptotic
flatness is possible since a necessary condition for ($a(r)\rightarrow 1$, $b(r)\rightarrow 1$, $c(r)\rightarrow 1$) is
\be
\mu _0 -3\mu _1 +3\mu _2 -\mu _3 =0
\ee
which is satisfied identically.

The simplest assumption of the nature of the perfect fluid is to use a polytropic equation of state that we write as
\be
P_\bot = P_r = P = P_{*} (\rho/P_{*})^\gamma
\ee
where $P_{*}$ and $\gamma >1$ are parameters  that characterize the matter. In order to study solutions, we first
change to dimensionless variables. The direct way to do that is to use $\ell = 1/\sqrt{\kappa P_{*}}$ as a length scale to
define $x=r/\ell$ and $\hat{m}=m\ell$ along with the obvious $\bar{\rho}=\rho/P_{*}$, $\bar{P}=P/P_{*}$
and also $\bar{f}_i (a,c) = f_i (a,c)/m^2$ and $\bar{g}_i (c) = g_i (c)/m^2$, which are still given by
(\ref{f-fctn})-(\ref{fg-funct}) with the replacements  $\mu_i \rightarrow \bar{\mu_i}=\mu_i / m^2$. The field
equations become a three-parameter system whose solutions ($a(x)$, $\bar{\rho} (x)$ etc...) are determined by the central
value of the density $\bar{\rho} (0)$. Typically, any value of $\bar{\rho} (0)$ (in a certain domain) corresponds to a
localized solution whose (coordinate) radius $r_s$ is determined by $\bar{\rho} (x_s) = 0$, that is  $r_s = \ell x_s$.

However, we may turn this system into a boundary value problem which is advantageous from the point of view of
numerical solution of this coupled system of ordinary differential equations. This can be done if we use as a
dimensionless radial variable $z=r/r_s$ and define also
a dimensionless radial extension $\epsilon=2(m r_s)^2$.
Pretending that we know $r_s$ in advance, the surface of the localized solution is now at $z=z_s=1$
and we may write the field equations in the following way (using $' = d/dz$):
\begin{eqnarray} \nonumber
\frac{2 b b'}{z} +\frac{b^2-1}{z^2} -\frac{\epsilon}{2} \left (\bar{g}_3 (c)-
\bar{g}_2 (c) \frac{b\left(c-zc'\right)}{c^2}\right) +\varpi \bar{\rho}= 0\\ \nonumber
\frac{2 b^2 a'}{z a} + \frac{b^2-1}{z^2}
-\frac{\epsilon}{2} \left (\bar{g}_3 (c)-\frac{1}{a} \bar{g}_2 (c)\right) -\varpi \bar{\rho} ^\gamma= 0\\
b^2 \left(\frac{a''}{a}+\frac{a' b'}{ab}+\frac{a'}{z a}+\frac{b'}{z b}\right)
-\frac{\epsilon}{2} \left (\bar{f}_3 (a,c) -\bar{f}_2 (a,c)\frac{b \left(c-z c'\right)}{c^2}\right) -
\varpi \bar{\rho} ^\gamma= 0\\ \nonumber
\bar{g}_2 (c)\frac{z a'}{a}+\frac{2 (b-1)}{b} \bar{f}_2 (a,c) = 0 \\ \nonumber
\gamma\bar{\rho} '+ \left( \bar{\rho}  + (\bar{\rho})^{2-\gamma}\right)\frac{a'}{a} = 0
\label{EqsDimensionless}.
\end{eqnarray}
An additional parameter, $\varpi =  \kappa P_{*} r_s^2 =(r_s /\ell)^2$ appears in the equations, but now we expect that
for each $(\varpi,\epsilon,\bar{\mu_i})$ there will typically be a single solution ($a(z)$, $\bar{\rho} (z)$ etc...)
that satisfies the simple boundary conditions $\bar{\rho}' (0) = 0$ and $\bar{\rho} (1) = 0$.

The solutions can be characterized in terms of a few physical parameters, namely the mass $M$ and size $r_s$ of the solutions.
We may define also the ``inertial mass'' $M_{I}$ and the physical radius given by
\be
 GM_{I} = G\int_{r\le r_s} d^3 x \sqrt{|g|} T_0^0 = \frac{r_s}{2} \bar{M}_{I} \varpi = \frac{\ell}{2} \bar{M}_{I} \varpi^{3/2}
  \ \ ,
 \ \ r_{phys} = \int_0^{r_s} \frac{dr}{b(r)} =  \ell z_{phys}  \varpi^{1/2}
\ee
where we use the dimensionless quantities
\be
      \bar{M}_{I} = \int_0^1 dz  \frac{z^2 a(z) \bar{\rho}(z)}{b(z)} \ \ , \ \ z_{phys} = \int_0^1 \frac{dz}{b(z)}
.\ee
Occasionally we will use the Schwarzschild radius $r_H=2GM$ in order to present the mass and to compare with GR results. We will calculate also the Vainshtein radius given for this theory by \cite{Hinterbichler}
\be
r_V = (8\pi GM/m^2 )^{1/3}
\label{RVainsht}
\ee
where we use the ``gravitational mass'' $M$ which is anyhow quite close to $M_{I}$. The way to obtain $M$ is by matching our solutions to the asymptotic vacuum behavior given in (\ref{yukawa}).

We will see that for the parameter space that we study, the Vainshtein radius is not large enough in order for the
Vainshtein mechanism to be realized.

\section{Numerical Techniques and Vacuum Solutions}
As mentioned already, it seems that no explicit solution to the above system of equations can be obtained,
therefore we have used a numerical technique to construct the solutions.
The radius $z_s$ and the parameters appearing in the mass term have, of course, to be fixed
 a priori. Without loosing generality, we can fix the scale and set $z_s=1$. We need also to specify the mass parameters
and we chose the representative values used already by Gruzinov and Mirbabayi \cite{Gruzinov2011} $\bar{\mu}_1 = 2$
$\bar{\mu}_2 = 3$ and $\bar{\mu}_3 = 5$. Note however that the solutions of Ref. \cite{Gruzinov2011} correspond
to a source of incompressible fluid, with a much smaller value of  $\epsilon$ than we use here.

The field equations were solved by employing a collocation method for boundary-value ordinary
differential equations, equipped with an adaptive mesh selection procedure \cite{colsys}.
The most efficient way we found to construct solutions consists
of  solving the equations in two steps that we now describe~:
\begin{figure}[b!]
\centering
\epsfysize=6cm
\mbox{\epsffile{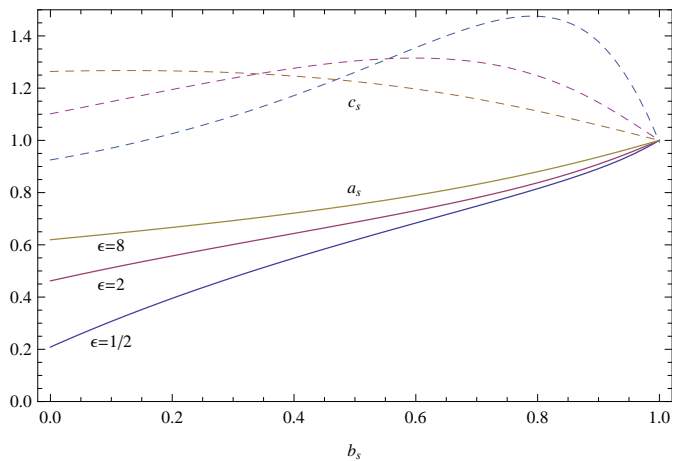}}
\caption{\label{vacuum_R_1}
Values of $a_s$ (lower curves) and of $c_s$ (upper dashed) as function of $b_s$ for the vacuum solution with $\epsilon=1/2$, $2$, $8$. The GR solutions correspond to the lines $a_s=b_s$ and $c_s=1$.}
\end{figure}
\begin{itemize}
\item First we solve the vacuum equations on an interval $z\in [z_s,\infty]$
by imposing the boundary conditions
\be
      c(z_s) = c_s \ \ , \ \ b(\infty) = 1
\ee
where $c_s$ is a constant and where the second condition ensures the solution to be
asymptotically flat.
The values $b_s \equiv b(z_s)$ and $a_s \equiv a(z_s)$ of the metric fields corresponding to $c_s$
can then be extracted from the numerical solution. Actually, some $c_s$ values may correspond to more than
a single vacuum solution, so in Fig. \ref{vacuum_R_1} we plot  $a_s$ and $c_s$ as a function of $b_s$.
The plot is for $\epsilon=1/2$, $\epsilon=2$ and $\epsilon=8$. Not shown in the plot are the GR curves,
$a_s =b_s$ and $c_s = 1$.
More generally, we expect such
families of asymptotically flat solutions (labelled e.g. by $b_s$) to exist
for generic values of the mass parameters $\mu_i$.
\item For the second step, i.e. solving for $0 \le z \le z_s$, it turns out essential to impose four boundary conditions
for our numerical methods to work efficiently, namely
\be
\label{bc_condition}
     b(0) = 1 \ \ , \ \ \bar{\rho}(0) = \bar{\rho}_0 \ \ ,  \ \ \bar{\rho}(z_s)=0 \ \ , \ \ c(z_s) = c_s
\ee
where $\bar{\rho}_0$ is a positive constant.
The  first condition is necessary for the metric to be regular at the center.
Two conditions on $\rho(r)$ turn out to be necessary, otherwise
 the linear equation of the density function leads to the trivial solution $\rho(r)=0$.
The last condition is for the continuity of the field $c(z)$ at $z=z_s$.
Since the system consists of three first order equations, it can accommodate only three boundary
conditions.
In order to be able to impose (\ref{bc_condition}),
the issue consists in extending the system to four
equations by supplementing  the equation $d \varpi / dr = 0$. The parameter $\varpi$ is therefore not an
input in this procedure, but is a part of the solution, thus fixing the radial size of the solution.
To proceed, we first choose a couple of values $(b_s,c_s)$ determined in the first step
 and then solve the system with the conditions (\ref{bc_condition}).
The value $\bar{\rho}_0$  has to be
fine tuned in such a way that $b(z_s)=b_s$. The fact that the equations for $b(r),c(r)$ are first order
then guarantee the metric functions to be  smooth at $z=z_s$.
\end{itemize}

\section{Solutions in Massive Gravity}
The numerical integration of the equations turns out quite involved and we did not address the
system throughout the full range of parameter values.
We discuss here the solutions for the polytropic power $\gamma = 2$ and
two values of the parameter $\epsilon$ which reflect the main features
of the solutions. We consider first a case where the radial extension parameter is large: $\epsilon = 8$.
 Then we report the results obtained in a case where
the radial parameter is intermediate, i.e. $\epsilon = 2$. Surprisingly, our numerical technique
works better for large $\epsilon$ and the pattern of solutions is obtained more easily in
this case.  As mentioned above, we set $z_s=1$ by an appropriate scaling.

\begin{figure}[b!]
\centering
\epsfysize=6cm
\mbox{\epsffile{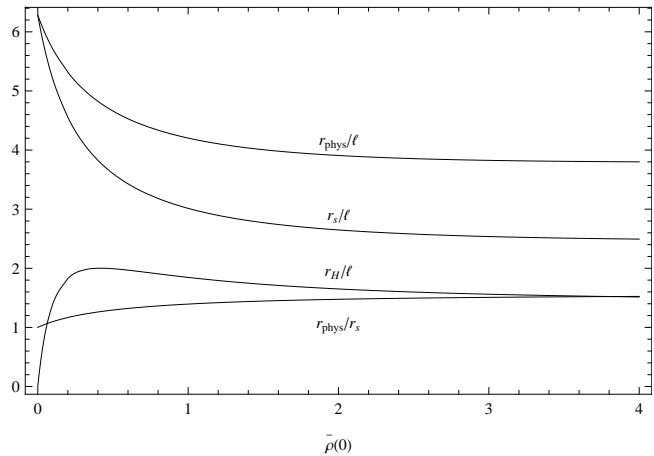}}
\caption{\label{RadialGR1}
Central density dependence of parameters of the GR solutions:  $r_{phys}$, $r_s$, $r_H =2GM$
and $z_{phys}=r_{phys}/r_s$.}
\end{figure}

\subsection{Solutions in General Relativity}
In order to test our numerical procedure and for the sake of comparison of the new results,
we first solved the equations in the case of GR.
In this case the independent fields are $a(r),b(r)$ and $\rho(r)$
and the corresponding (first order) equations have to be solved with the following
boundary conditions~:
\be
       b(0)=1 \ \ , \rho(r_s)=0 \ \ , \ \ a(r_s)=b(r_s)
\ee

By the rescaling explained above $r_s$ will correspond to $z=r/r_s=1$, i.e. $z_s =1$
and the only free parameter is $\varpi$.
The solutions can then be constructed for different values of  $\varpi$.
Our numerical results show that the solutions exist for values of $\varpi$ in a finite interval,
i.e. for $\varpi_m < \varpi < \varpi_M$ (we find $\varpi_m \approx 6$ and $\varpi_M \approx 40$)
with a single solution for each $\varpi$.
 In the limit  $\varpi \to \varpi_M$, the density function approaches $\rho(r)=0$
and vacuum solution (i.e. Minkowski space-time) is recovered.
In the limit  $\varpi \to \varpi_m$, we observe that the value of metric component $g_{00}$ at the center,
i.e. $a(0)$ tends to zero and the solution becomes singular. The values of density at the center $\rho(0)$
diverges, but the mass stays finite. Fig \ref{RadialGR1} shows the behavior of the main parameters of the solutions as functions of
the central density $\bar{\rho}(0)=\rho(0)/P_{*}$.

\begin{figure}[b!]
\hbox to\linewidth{\hss%
    \resizebox{7.6cm}{6.2cm}{\includegraphics{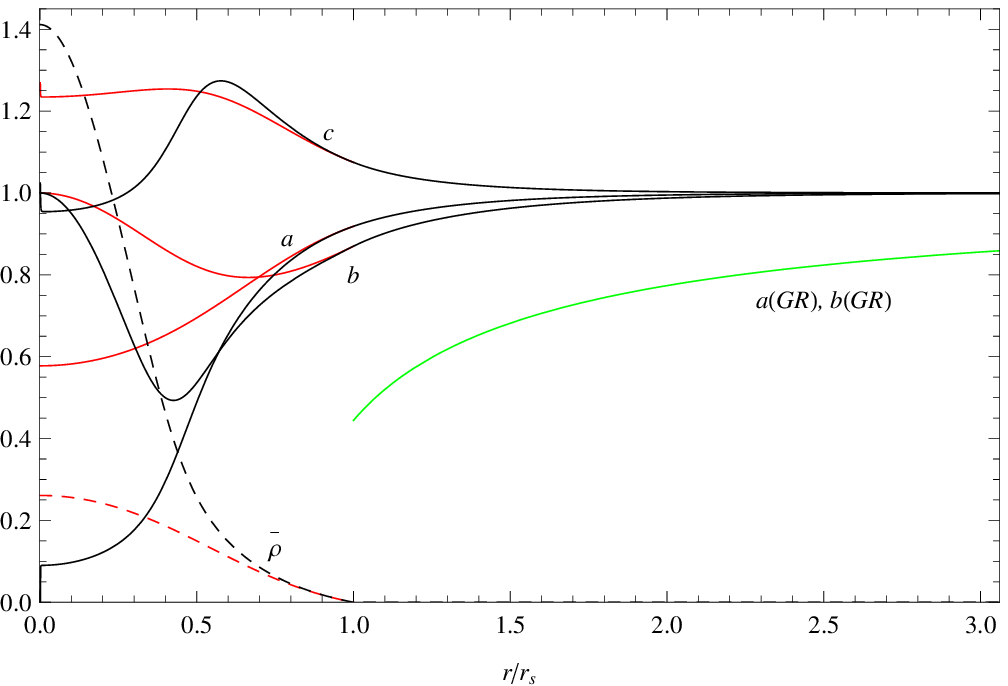}}
\hspace{5mm}%
        \resizebox{7.6cm}{6.2cm}{\includegraphics{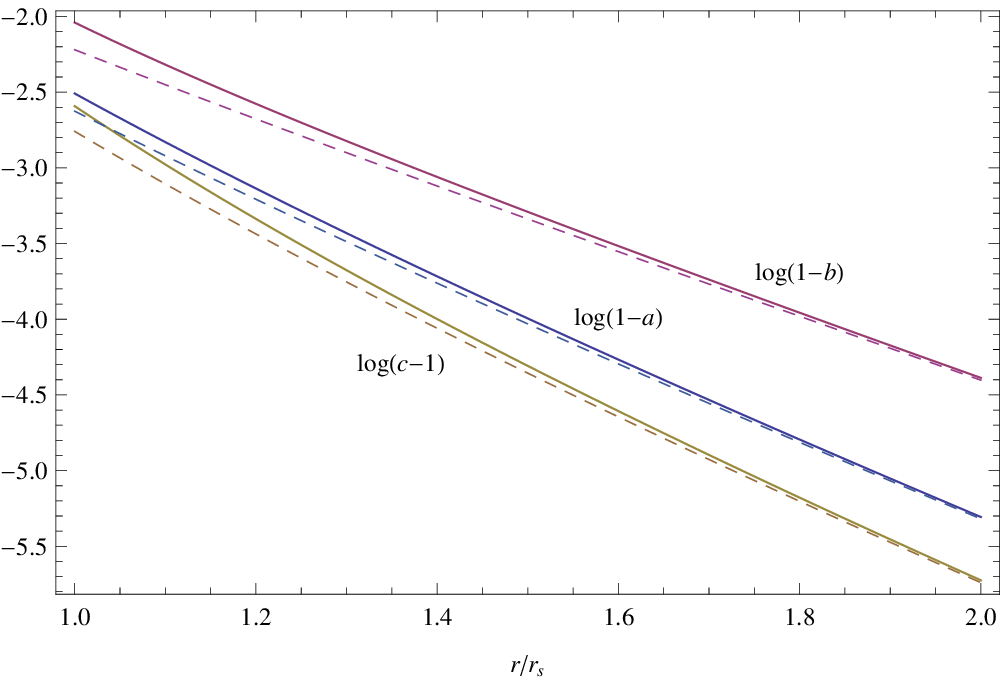}}
\hss}
    \caption{
Profiles of two solutions with the same exterior behavior corresponding to $\epsilon=8$, $b_s=0.870$, $c_s = 1.075$:
Left: The curves for $\bar{\rho}(0)=0.26$ (red) and $\bar{\rho}(0)=1.41$ (black). Added for comparison is the exterior $a(r)=b(r)$ curve of GR with the same mass and $r_s$ as the second solution: $GM/\ell=2.39$, $r_s/\ell=5.96$. Right: Fit of the exterior solutions
(solid lines) by the asymptotic solutions (\ref{yukawa}) (dashed lines). Beyond $r/r_s=2$ the difference between the
solutions cannot be resolved in the figure.}
\label{profile}
\end{figure}


\subsection{Massive Gravity Solutions: Case $\epsilon = 8$}

\begin{figure}[t!]
\hbox to\linewidth{\hss%
 \resizebox{8.0cm}{6.0cm}{\includegraphics{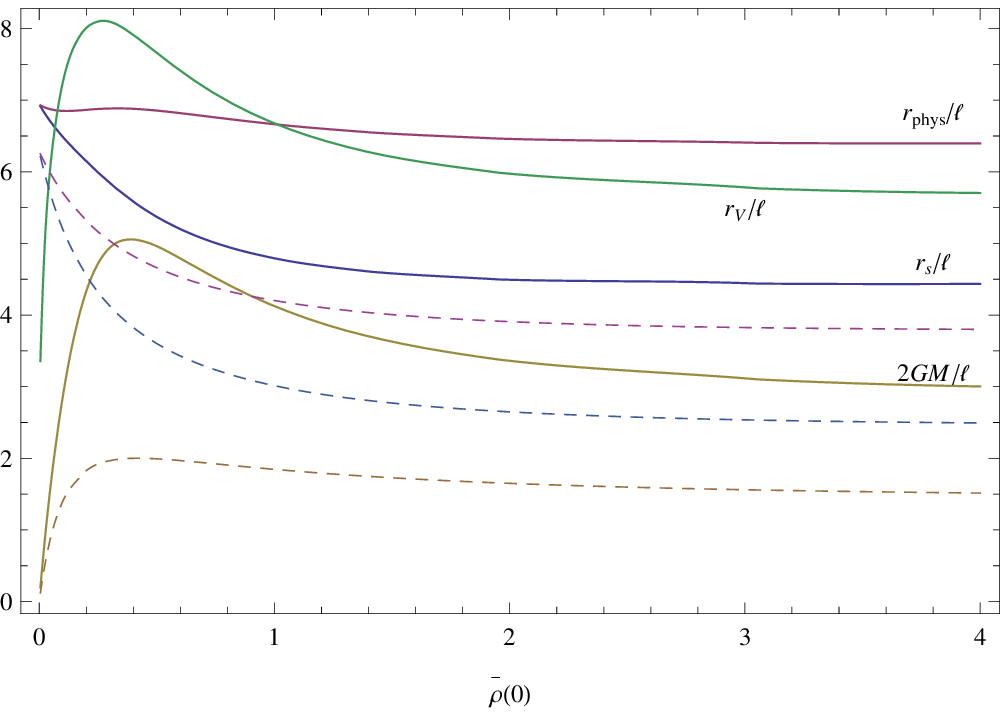}}
  \hspace{5mm}%
      \resizebox{3.6cm}{6.0cm}{\includegraphics{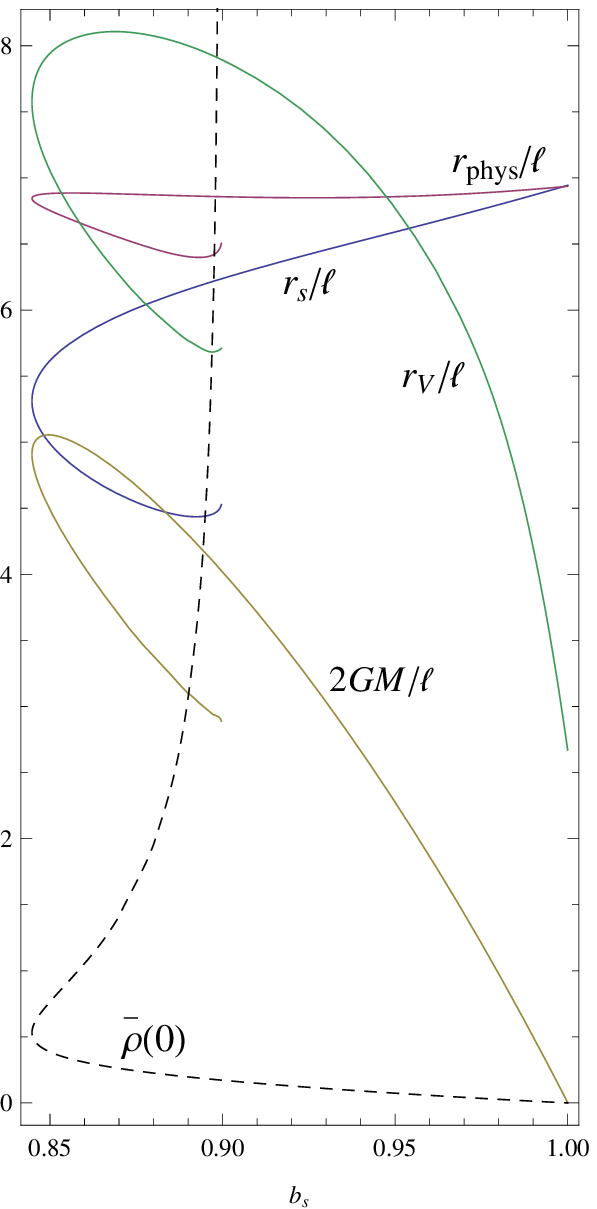}}
\hss}
    \caption{ Characteristics of the solutions for $\epsilon=8$. Left: Mass and radii of the solutions
    as functions of the density at the origin $\rho(0)$: $M$, $r_s$,  $r_V$ and $r_{phys}$ . The dashed curves are the corresponding GR ones. Right: The same parameters plus $\bar{\rho}(0)$ vs $b_s$. }
\label{physical8}
\end{figure}

Next we move to the main subject of this work, the MGR solutions. Our results are summarized in Fig. \ref{physical8} where several parameters characterizing the solutions are plotted. The main structure is seen from Fig. \ref{physical8} -left. The right hand part shows further details.

One new feature is that the number of values of $\rho(0)$ leading to a matching of the metric functions
at the surface, i.e. to $b(z_s)=b_s, c(z_s)=c_s$
depend very sensitively on the values of $\epsilon$ and of $b_s$ or $c_s$. Actually, only the small
rightmost interval of Fig. \ref{vacuum_R_1} is realized in the present case:
For $b_s < 0.845$ (which correspond to $c_s > 1.089$), the values $b(z_s)$ computed from the interior solutions
 are always larger than $b_s$ and no continuous solution is available.
In the interval $ 0.845 < b_s < 0.899$ ($ 1.058 < c_s < 1.089$) there are two solutions corresponding to the same values of $(b_s,c_s)$, they are distinguished by the values of $\rho(0)$. That is two different interior solutions correspond
to the same exterior solution.
Profiles of  two such solutions for ($b_s = 0.870$, $c_s = 1.075$) are  illustrated in Fig. \ref{profile} (left part). The two solutions have ($\bar{\rho}(0) \approx 1.41$, $\varpi \approx 21.2$, $GM/\ell \approx 1.85$)
and ($\bar{\rho}(0) \approx 0.26$, $\varpi \approx 35.5$, $GM/\ell \approx 2.39$). The right-hand part of this figure
demonstrate the very good fit of the exterior solution by the asymptotic solution (\ref{yukawa}) starting already around $r/r_s = 1.5$. This figure shows the domain where the actual and perturbative solutions are still discernable.

The existence of two branches occurs also in
GR as can be inferred from the GR plots (Fig. \ref{RadialGR1}), but it is more pronounced in MGR.
At any rate, this does not
violate any basic principle, since the two different interior solutions are actually solutions of two systems
differing by the values of $\varpi$. From a physical point of view, these two solutions have different sizes,
so it is only when they are written in terms of the ratio $r/r_s$, that the exterior solutions match.

The two branches of solutions do not persist beyond $b_s =0.899$ (or $c_s = 1.058$). Indeed we observe that the value
$a(0)$ associated with one of the branches decreases and becomes null in the limit $b_s = 0.899$. This is of course the
limit $\rho(0)\to \infty$ mentioned above - see Fig. \ref{physical8} right. Therefore only one solution exists for any
$ 0.899 \le b_s < 1$ (or  $1 < c_s \le 1.058$).
In the limit ($b_s \to 1$, $c_s \to 1$), the matter density converges uniformly to $\rho(r)=0$ and the limiting
configuration is just the vacuum.

As for the parameter $\varpi=(r_s/\ell)^2$ defined above, we find that only a subset of the values  of the parameter
$\varpi$ allowing for polytrope solutions in GR lead also to MGR solutions. Also there is a small interval of $\varpi$ (for instance $19.8 < \varpi  < 20.4$ which corresponds to $\rho(0)\approx 4.5$) for which two solutions exist
 -- see the region near the local minimum of the $r_s/\ell$ curve in Fig. \ref{physical8} -right. In other words,
 in a certain domain in the parameter space two values of the central density correspond to the same value of $r_s$.
 The plot of $r_s/\ell$ vs. $\rho(0)$ develops small oscillations which cannot be resolved in Fig. \ref{physical8} -left.
 This feature seems peculiar to MGR.
 As another feature, we observe that, in the critical limit where $a(0)$ tends to zero,
 $\rho(0)$ diverges while the inertial mass remain finite as in the GR case.

Last we turn our attention to the Vainshtein radius $r_V$. As is obvious from Fig. \ref{physical8}, its maximum value is around $r_V \approx 8$ which is never significantly larger than $r_s$. We therefore do not expect the Vainshtein mechanism to operate for this family of solutions. On the other hand, the perturbative (weak field) solution - Eq. (\ref{yukawa}) approximates very well the solutions as is demonstrated in Fig. \ref{profile} -right.

\subsection{Massive Gravity Solutions: Case $\epsilon = 2$}

\begin{figure}[t!]
\hbox to\linewidth{\hss%
 \resizebox{8.0cm}{6.0cm}{\includegraphics{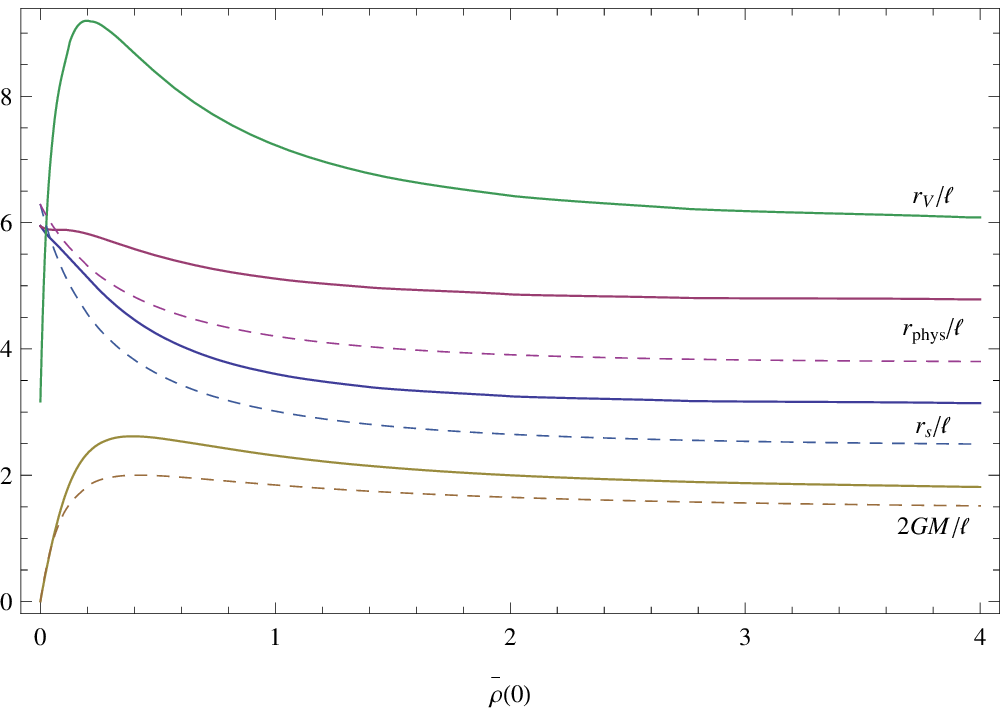}}
  \hspace{5mm}%
      \resizebox{3.6cm}{6.0cm}{\includegraphics{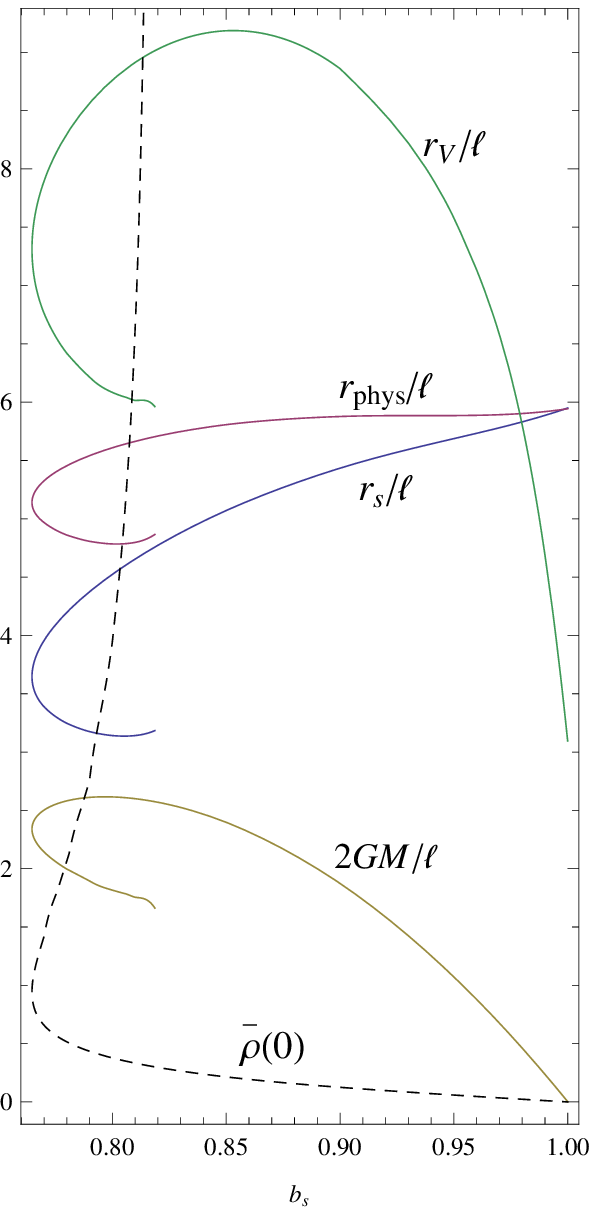}}
\hss}
    \caption{ Characteristics of the solutions for $\epsilon=2$. Left: Mass and radii of the solutions
    as functions of the density at the origin $\rho(0)$: $M$, $r_s$,  $r_V$ and $r_{phys}$ . The dashed curves are the corresponding GR ones. Right: The same parameters plus $\bar{\rho}(0)$ vs $b_s$. }
\label{physical2}
\end{figure}

Solutions in this case exist for a larger interval of values of $b_s$ or $c_s$. However, it became quite difficult
to construct them. The reasons for this difficulties is technical and
results from the elimination of the field $a(r)$ in the equations as explained at the end of Sec.
\ref{spherical}. Indeed, it turns out that when the parameter $c_s$ becomes large,
both numerator and denominator (denoted $X$ and $Y$ in (\ref{a_expression})) develop a zero
   at some common value of the radial variable, say $r=r_0$. This creates numerical difficulties
   although the resulting field $a(r)$ is continuous and differentiable.
  We checked that this problem also occurs
  when $b(r)$ is eliminated instead of $a(r)$ from  Eqs. (\ref{EqsNonCompact}). Avoiding this problem would likely require   to adopt a  different coordinate system but this direction was not taken in the present paper.
  Anyhow, the pattern of the solution is similar to the case $\epsilon = 8$ and the counterpart of Fig. \ref{physical8} for the case $\epsilon=2$ is included as Fig. \ref{physical2}. Solutions exist now in a somewhat larger interval of $0.765 \le b_s \le 1$ where $b_s = 0.818$
divides between the two branch domain and the one branch domain: for $0.765 < b_s < 0.818$ two interior solutions exist for a single exterior one, while for $ 0.818 \le b_s \le 1$ there exists a single branch.

Here too, the Vainshtein radius $r_V$ is not large enough for turning on the Vainshtein mechanism. As is obvious from Fig. \ref{physical2}, its maximum value is around $r_V \approx 9$ which is never significantly larger than $r_s$. On the other hand, the perturbative (weak field) solution - Eq. (\ref{yukawa}) still approximates very well the solutions very much as
demonstrated in Fig. \ref{profile} (right part).

\subsection{ Other Cases }

\begin{figure}[t!]
\hbox to\linewidth{\hss%
 \resizebox{7.6cm}{6.0cm}{\includegraphics{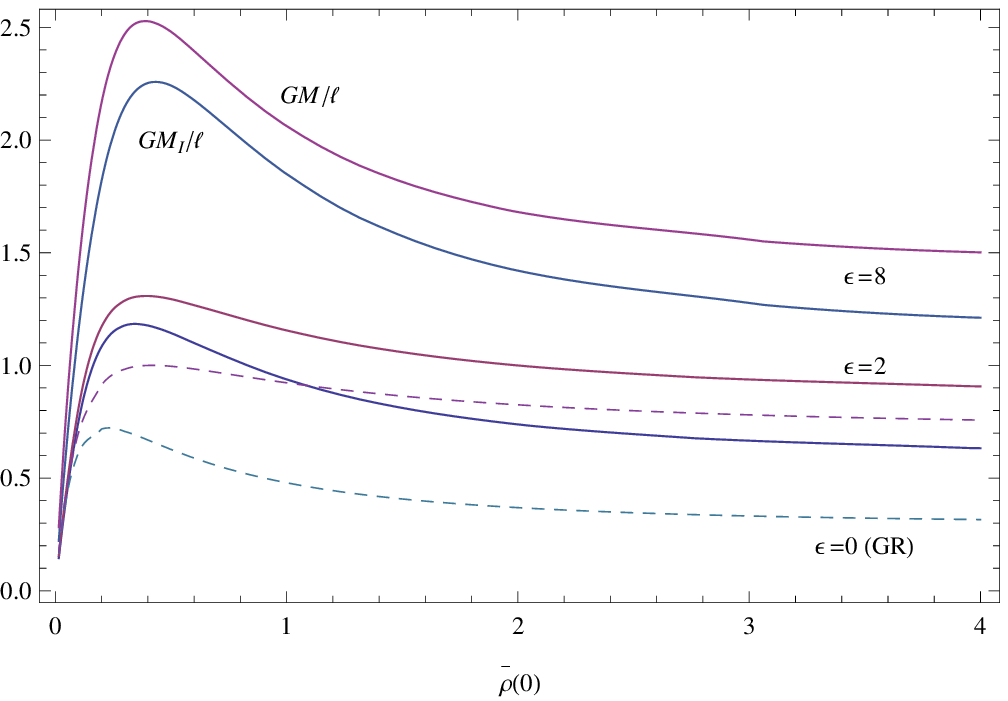}}
  \hspace{5mm}%
      \resizebox{7.6cm}{6.0cm}{\includegraphics{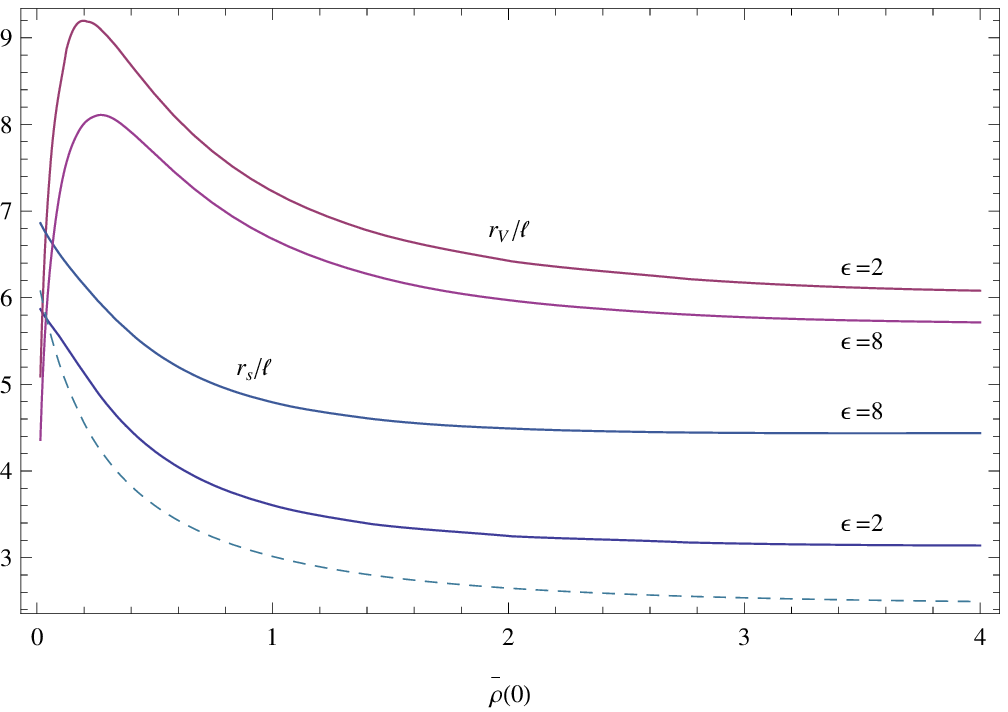}}
\hss}
    \caption{ Characteristics of the solutions for  $\epsilon =2$, $8$ together with the GR curves (dashed). Left: The mass parameters $M_I$ (lower blue lines) and $M$ (upper red lines) as a function of the central
density. Right: The radial parameters $r_s$ (lower blue) and $r_V$ (upper red). }
\label{MassesRadii}
\end{figure}

The $\rho(0)$-dependence of the mass and radial parameters are shown for these two values of $\epsilon$
together with the GR curves in Fig. \ref{MassesRadii}.

We further studied the system also for smaller values of $\epsilon$
and obtained results qualitatively similar to the case $\epsilon = 2$.
In particular the Mass parameters $M$, $M_I$ approach continuously their corresponding
GR values. In all cases we find a similar behavior of a maximal
mass for a certain central density beyond which the mases decrease and the solutions are unstable.
However, it seems that the graviton mass term increases considerably the mass range of possible solutions. The Vainshtein radius increases as $\epsilon$ decreases and we expect this tendency to continue
such that the Vainshtein mechanism will take action and a Newtonian domain will emerge at  intermediate
distances from the source.

 Finally, we studied  the equations in  the case when only the quadratic mass term is
present, i.e. $\sigma_3 = \sigma_4 = 0$.
Preliminary results obtained in this case confirm that solutions exist roughly with the same pattern
as in the previous cases that were studied  more extensively. However, the construction also reveals that
 the numerical difficulties related to the occurrence of zeroes in the numerator
and denominators of $a(r)$ in Eq. (\ref{a_expression}) are  more severe than in the case where the
higher order terms are present.

\section{Conclusion}

This paper summarizes the first steps of analysis of spherically-symmetric solutions in a recently proposed
ghost-free model for massive gravity (MGR), where the non-linear effects cannot be treated perturbatively.

We have constructed  new families of spherically-symmetric solutions in MGR generated by perfect fluid
matter sources. The solutions are obtained non-perturbatively by
solving the underlying non-linear equations, they are regular over the whole space-time,
concentrated inside a sphere of radius $r_s$, and characterized by an energy density and a pressure
related by a power law: a polytrope. The equations are a system of coupled non-linear equations
labeled by two physical parameters: the graviton mass (or inverse range) $m$ and the length scale $\ell$ defined in terms of the energy density scale $P_*$ by $\ell =  1/\sqrt{8 \pi G P_*}$ which enters naturally into the equations.

Within the range of parameters for which  we have been able to construct numerically reliable solutions, we solved the equations along curves in parameter space of constant $m r_s$. The solutions were obtained for a rather limited domain  where $m r_s$ is of order one. We believe that this limitation of our solutions is just a technical
problem which can be overcome by  more sophisticated solving techniques. This is of paramount importance
since a more extensive analysis is evidently required in order to go beyond the $\epsilon \sim 1$
range.

We found that the solutions exist only for values of the ratio $r_s/ \ell$ limited to quite a small interval. This is the reason why the Vainsthein radius turns out to be not much larger than $r_s$ in the region of parameter space we mapped, and the Vainshtein mechanism is not operative. In other words, the limit of vanishingly small $m$ is out of reach for the parameters we used.

For a given central density  $\bar{\rho}(0)$ the mass and size of the solutions are larger than the
corresponding GR ones and increase with $\epsilon$. The  Vainsthein radius on the other hand decreses with $\epsilon$
pointing to the possibility that the Vainshtein mechanism turns on for small enough values of $\epsilon$.

As mentioned already, studying this system for small enough  $\epsilon$ is a major part of this program which is currently in progress. Also interesting is a systematic survey of solutions for the full 2-dimensional parameter space of mass parameters $\sigma_i$. It is already known that the case ($\sigma_2=1$, $\sigma_3=-1$, $\sigma_4=1$)  is special in that the Vainshtein mechanism does not operate \cite{KoyamaEtAl2} since the system is ``not nonlinear enough''. It is expected that several more special points in this parameter space exist which may shed light on its structure and hint towards a preferred mass term for MGR.

Also the value $\gamma = 2$ of the polytropic power was chosen for convenience, and extending our work for $\gamma > 2$ as well as for $1< \gamma < 2$ is a self-evident direction for further study.

\end{document}